\documentclass[twocolumn,aip,showpacs,apl,tightenlines,amsmath,amssymb,superscriptaddress]{revtex4}
\usepackage{amssymb} 
\usepackage{graphicx} 

\usepackage{dcolumn} 
\usepackage{amsmath} 
\usepackage{bm}
 
\usepackage{colordvi}

\begin{document} 
 
\title{A scheme for spin transistor with extremely large on/off
  current ratio} 

\author{L. Wang} 
\affiliation{Hefei National Laboratory for Physical Sciences at 
Microscale and Department of Physics, 
University of Science and Technology of China, Hefei, 
Anhui, 230026, China} 
\author{K. Shen} 
\affiliation{Hefei National Laboratory for Physical Sciences at 
Microscale and Department of Physics, 
University of Science and Technology of China, Hefei, 
Anhui, 230026, China} 
\author{S. Y. Cho} 
\affiliation{Centre for Modern Physics and Department of Physics, 
  Chongqing University, Chongqing 400044, China} 
\author{M. W. Wu\footnote{Corresponding author; Electric address: mwwu@ustc.edu.cn}} 
\affiliation{Hefei National Laboratory for Physical Sciences at 
Microscale and Department of Physics, 
University of Science and Technology of China, Hefei, 
Anhui, 230026, China}

\date{\today} 
 
\begin{abstract} 
 Quantum wires with periodic local Rashba spin-orbit couplings are 
 proposed for a higher performance of spin field-effect transistor. 
 Fano-Rashba quantum interference due to the spin-dependent modulated structure 
 gives rise to a broad energy range of vanishingly small transmission. Tuning 
 Rashba spin-orbit couplings can provide  the on- or off-currents 
 with 
 extremely large on/off current ratios 
 even in the presence of a strong disorder. 
 \end{abstract} 
 

\maketitle 

\section{Introduction}

 Coherent manipulations of electron spin in nanoscale devices 
 have made it promising to realize spintronics as well as 
 quantum information processing and computation \cite{development}. 
 In 1990, especially, Datta and Das proposed a spin transistor in which electron spin 
 can be manipulated by varying 
 electric fields via spin-orbit couplings (SOCs) \cite{Datta and Das}. 
 Spin transport have been intensively investigated in various types of such spin 
 field-effect transistors (SFETs) 
 in associations with  spin filters or polarizers, spin currents, 
 spin valves, and so on. However, the device working performances have not been 
 studied yet much. 
 Even the device performance rates in SFETs seem to be much lower than 
 in charge field-effect transistors (FETs). 
 Actually, for instance, the on/off current 
 ratio is smaller than $10^3$  in a SFET with
 $T$-shaped structure \cite{T} as well as that in a dual-gate SFET \cite{wan}. 
  On the contrary, charge field-effect transistors 
 have reached a higher performance rate. 
 Indeed, recent high-mobility FETs 
 using ZnO nanorods \cite{ratio1}, 
 poly(3-hexylthiophene) thin film \cite{ratio2}, 
 carbon nanotubes \cite{ratio3}, and 
 p-GaN/u-$\mathrm{ Al_xGa_{1-x}}$N/u-GaN junction heterostructure \cite{ratio4} 
 have been demonstrated with $10^4 \sim 10^6$ on/off current ratios. 
 Since spin coherence can be more fragile than charge 
 decoherence, 
 device disorders may also spoil the performance 
 rate more significantly in SFETs than in charge field-effect transistors. 
 
 In this paper, we propose a SFET with high performance rate. 
 To do this, we introduce a quantum wire with Rashba SOC 
  modulations \cite{Rashba strength}. A local Rashba SOC 
 in a quantum wire 
 gives rise to a spin-dependent Fano effect that is the quantum 
 interference of 
 electron propagating through both 
 the continuum energy channel and a localized electronic state resulting from 
 the SOC. 
 Due to the Rashba-Fano effect, electron transmission has 
 asymmetric antiresonance  dips \cite{wire,fano shapes,wire1}. 
 Further, the periodic SOC modulation makes the 
 energy range of vanishingly small electron transmission 
 broader. 
 This allows us a large on/off current ratio 
 by tuning the SOC modulations. 
 Even for a strong device disorder, 
 the proposed SFET is shown to have higher performance rate 
 with larger on/off current ratio than about $10^5$. 

\begin{figure}
\centering
\includegraphics[width=7.5cm]{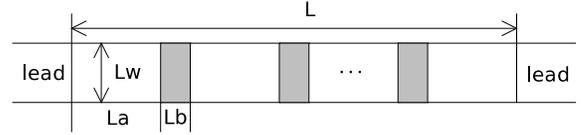}
\caption{ Schematic view of quantum wire with periodic local Rashba
  couplings. The width and length of the quantum wire are $L_w$ and $L$, respectively.
  The gray areas represent each Rashba region with its length $L_b$.
  The length of normal regions without SOC is $L_a$.
  In the text, an integer $N$ denotes the number of Rashba regions. }
\label{fig1}
\end{figure}

\section{Model and Formalism}

 A quantum wire under consideration is shown in Fig.~1. 
 The wire of length $L$  is connected to two ideal leads. Its width is 
 $L_w$, which determines the transverse propagating channels. 
 The Rashba SOC is controlled in the gray regions 
 with its length $L_b$, while the normal regions without the 
 SOC has their size $L_a$. 
 The number of Rashba  regions is denoted by $N$. 
 We apply a weak uniform perpendicular magnetic field, which 
 guarantees  that Landau levels 
are neglected, in the whole device including the leads. 
 The Zeeman effect splits the on-site electronic energy 
 into $V_{\sigma}={\sigma}V_0$ with $\sigma=\pm$ for 
 spin-up $(\uparrow)$ and -down $(\downarrow)$. 
 To describe the quantum wire, then we introduce a tight-binding 
 Hamiltonian with nearest-neighbor hopping: 
\begin{eqnarray} 
 H\!\!\!&=& 
 \!\!\!\sum_{lm\sigma}\epsilon_{l,m,\sigma}c^{\dagger}_{l,m,\sigma}c_{l,m,\sigma} 
  -t\sum_{lm\sigma}(c^{\dagger}_{l+1,m,\sigma}c_{l,m,\sigma} + 
  \mathrm{H.C.}) 
  \nonumber \\&& 
  -t\sum_{lm\sigma}(c^{\dagger}_{l,m+1,\sigma}c_{l,m,\sigma}+ 
  \mathrm{H.C.})+H_{R}\ , 
  \label{H} 
\end{eqnarray} 
 where $(l,m)$ represent the site in the space representation of 
 $(x,y)$. The on-site energy is $\epsilon_{lm\sigma} = 4t + \sigma 
 V_0$ with the hopping integral $t=\hbar^2/(2m^* a^2)$, where 
 $m^\ast$ and $a$ are the electron effective mass and lattice 
 constant, respectively. The Hamiltonian $H_R$ describing the Rashba 
 SOC modulation is given by 
\begin{eqnarray} 
  H_R&=&\lambda i \sum_{lm\sigma\sigma^\prime} 
  \left(c^{\dagger}_{l+1,m,\sigma}c_{l,m,{\sigma}'} 
 \sigma_y^{\sigma\sigma^\prime} 
 \right. 
 \nonumber \\ && \hspace*{1.6cm} 
 \left. -c^{\dagger}_{l,m+1,\sigma}c_{l,m,\sigma^\prime} 
 \sigma_x^{\sigma\sigma^\prime} 
 +\mathrm{H.C.}\right), 
\end{eqnarray} 
 where $\lambda=\alpha/2a$ is the Rashba SOC coefficient. 
 $\sigma_{x,y}$ are the Pauli matrices.

 \begin{figure}
\centering
\includegraphics[width=7.cm]{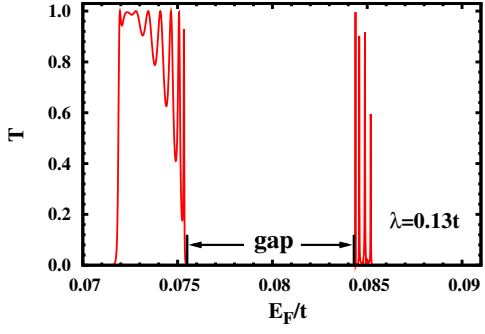}
\caption{(Color online) Transmittance $T$ as a function of the Fermi
 energy of the leads for $N=9$. The Rashba regions are chosen as
 $L_{b}=10a$ for the SOC strength $\lambda=0.13t$.
 Other parameters are $L_{a}=3L_b$ and $L_{w}=L_b$.
 The ``gap" indicates the energy region where the transmission amplitude
 is very small. Here, the gap is developed by the periodic Rashba SOC
modulations.}
\label{fig2}
\end{figure}

 Within the Landauer-B\"uttiker framework \cite{LB}, at zero 
 temperature,  the transport current is given by 
\begin{equation} 
 I = \frac{e}{h} \sum_{\sigma\sigma'} \int^{\mu_{2}}_{\mu_{1}} 
     T^{\sigma\sigma'}(\varepsilon) d\varepsilon, 
\end{equation} 
 where the spin-resolved transmission amplitude \cite{LB} is 
\begin{equation} 
 T^{\sigma\sigma^\prime}(\varepsilon) 
 = 
 \mbox{Tr}\left[\Gamma_1^\sigma \ 
 {\cal G}_{1M}^{\sigma\sigma^\prime(r)}(\varepsilon) \ 
 \Gamma_M^{\sigma^\prime} 
 \ {\cal G}_{M1}^{\sigma^\prime\sigma(a)}(\varepsilon)\right], 
\end{equation} 
 with the tunnel couplings $\Gamma_{1(M)}$ between the wire and the leads. 
 ${\cal G}_{1M}^{\sigma\sigma^\prime(r)}$ and 
 ${\cal G}_{M1}^{\sigma^\prime\sigma(a)}$ are the retarded and advanced 
 Green's functions \cite{greenfunction}, respectively. 
 Here, $\mu_{1}$ ($\mu_{2}$) is the Fermi energy of the 
   left (right) lead. 
 
 The energies of the transverse propagating channels are given by 
 $\varepsilon_n=2t\{1-\cos[n\pi/(L_w/a+1)]\}$ with the channel index $n$. 
 We consider the lowest energy channel $n=1$. In our numerical 
 calculation, the lowest energy is $\epsilon_1\simeq0.081t$ for $L_w=10a$. 
 It is assumed that only spin-down 
 electrons can propagate into or out of the wire structure. 
 To make it sure, the Fermi energy $E_F$ can be adjusted as the energy 
 between $\varepsilon_1-V_0$ and $\varepsilon_1+V_0$. 
 Thus, 
 only $T^{\downarrow\downarrow}$ is responsible for electron transport, i.e., 
 the total transmission amplitude becomes $T = T^{\downarrow\downarrow}$. 

\section{Results} 

 In Fig.~2, 
 we plot the transmission amplitude as a function of Fermi energy 
 for $N=9$ and $\lambda=0.13t$. Numerical parameters were chosen as $V_0=0.01t$, 
 $L_{b}=10a$, $L_{a}=3L_b$ and $L_{w}=L_b$. Note that due to the serial SOC 
 modulations the electron transmission 
 is vanishingly small in a wide range of energy. 
 Roughly, the energy range is from $E_F=0.076t$ to $E_F=0.084t$. 
 This can play a significant role for the device 
 performance, i.e., particularly the off current. 
 To show clearly the effects of periodic Rashba SOC modulations, 
 in Fig. 3 (a), the transmission amplitudes are plotted for 
 $N=1,3,5,7,9,$ and $12$. It is shown that the gap where the 
 vanishingly small transmission happens becomes wider and deeper as the number 
 of Rashba SOC regions increases. 
 Also, inside the gap, Fano-Rashba antiresonances appear to give a 
 much smaller transmission through the quantum wire, which can provide smaller 
 off currents. 
 Further, as the SOC strength decreases, one find a high 
 transmission within the gap. In Fig. 3 (b), 
 the high transmission for $\lambda = 0.08t$ is shown in the energy range 
 of low transmission for $\lambda = 0.13t$. 
 The high transmission in the gap does not change very much as 
 the number of SOC regions increases. 
 Then, this may provide a high current as an on current. 
 
 \begin{figure}
\centering
\includegraphics[width=7.cm]{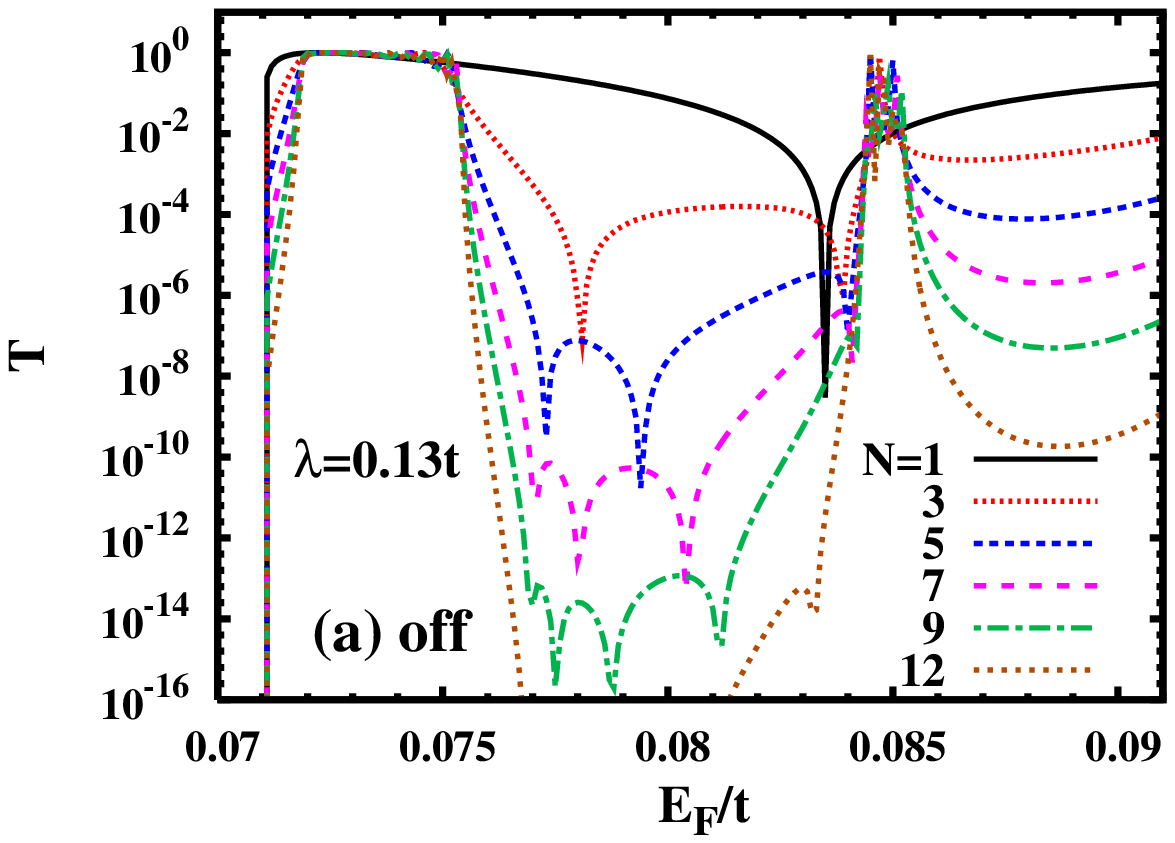}
\includegraphics[width=7.cm]{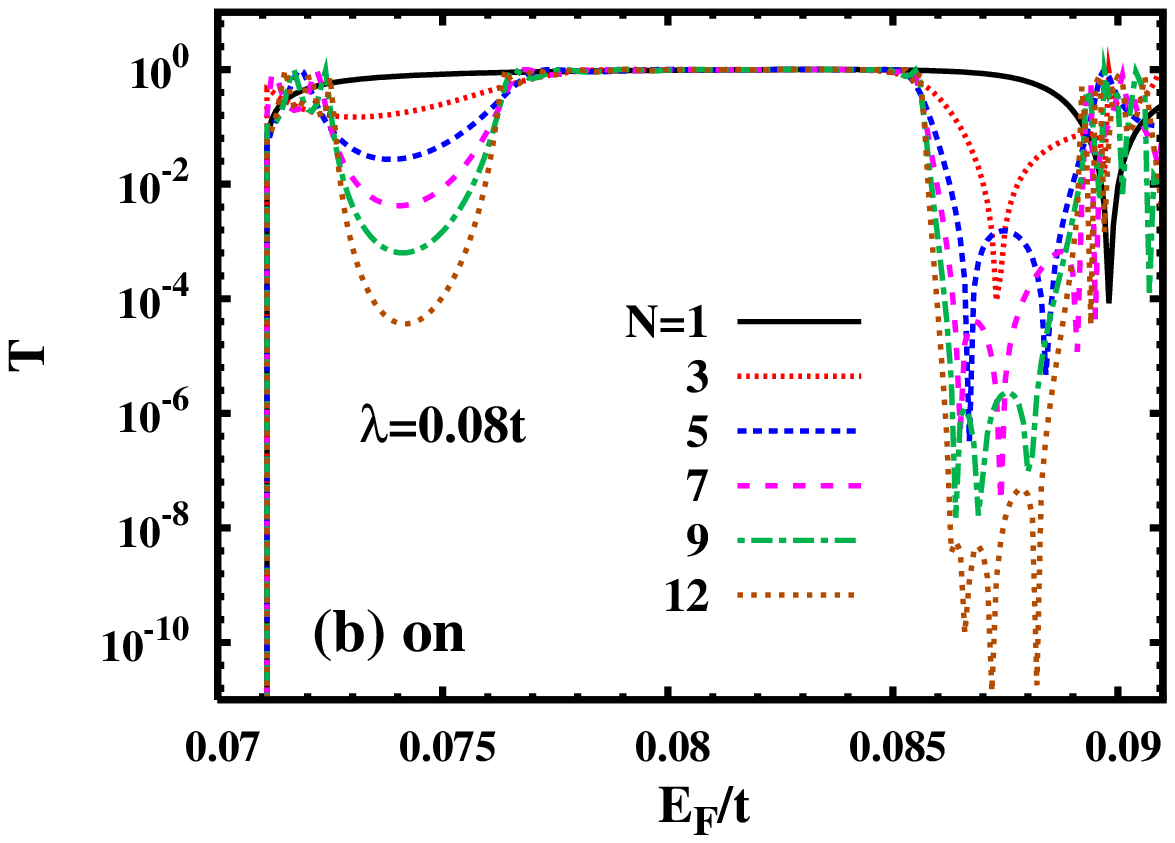}
\caption{(Color online)
 Transmittances $T$ as a function of the
Fermi
 energy of the leads for various SOC modulations.
 The Rashba regions are chosen as
 $L_{b}=10a$ for the SOC strength (a) $\lambda=0.13t$ (off current) and (b)
  $\lambda=0.08t$ (on current).
 Other parameters are $L_{a}=3L_b$ and $L_{w}=L_b$. }
\label{fig3}
\end{figure}
 
 In Fig.~4, the spin transport current is plotted as a function 
 of the SOC strength $\lambda$ for the voltage bias window 
 $[0.078t,0.082t]$. 
 It should be noted that there are three regimes of spin current as 
 the SOC varies, i.e., 
  (i) high current regime $( 0 \leq \lambda \lesssim 0.1 \ t)$, 
  (ii) transit current regime $( 0.1\ t \lesssim \lambda \lesssim 0.12\ t)$, 
 and (iii) low current regime $( 0.12\ t \lesssim \lambda)$. 
 For instance, from the numerical calculation, 
 we obtain the on/off current ratios: $1 \times 10^{4}$ for $N=3$, 
 $2.7 \times 10^{12}$ for $N=9$, and $6.8 \times 10^{15}$ for $N=12$. 
 Then, our SFETs show the high device performance with the on/off 
 current ratio bigger than that of  EFTs in 
Refs.~\onlinecite{ratio1,ratio2,ratio3,ratio4} and \onlinecite{ratio5}. 
 
 \begin{figure}
\centering
\includegraphics[width=7cm]{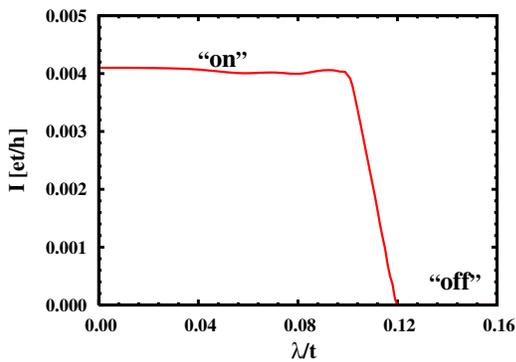}
\caption{ (Color online) Spin transport current
 as a function of the Rashba SOC strength
 for $N=9$.
 The voltage bias window is chosen as
 $[0.078t,0.082t]$ within the gap in Fig. \ref{fig2}.
 Other parameters are $L_{b}=10a$, $L_{a}=3L_b$ and $L_{w}=L_b$.
 The ``on" and ``off" indicate the high and low currents
 for the device performance rate.}
\label{fig4}
\end{figure}
 
\begin{figure}
\centering
\includegraphics[width=4.2cm]{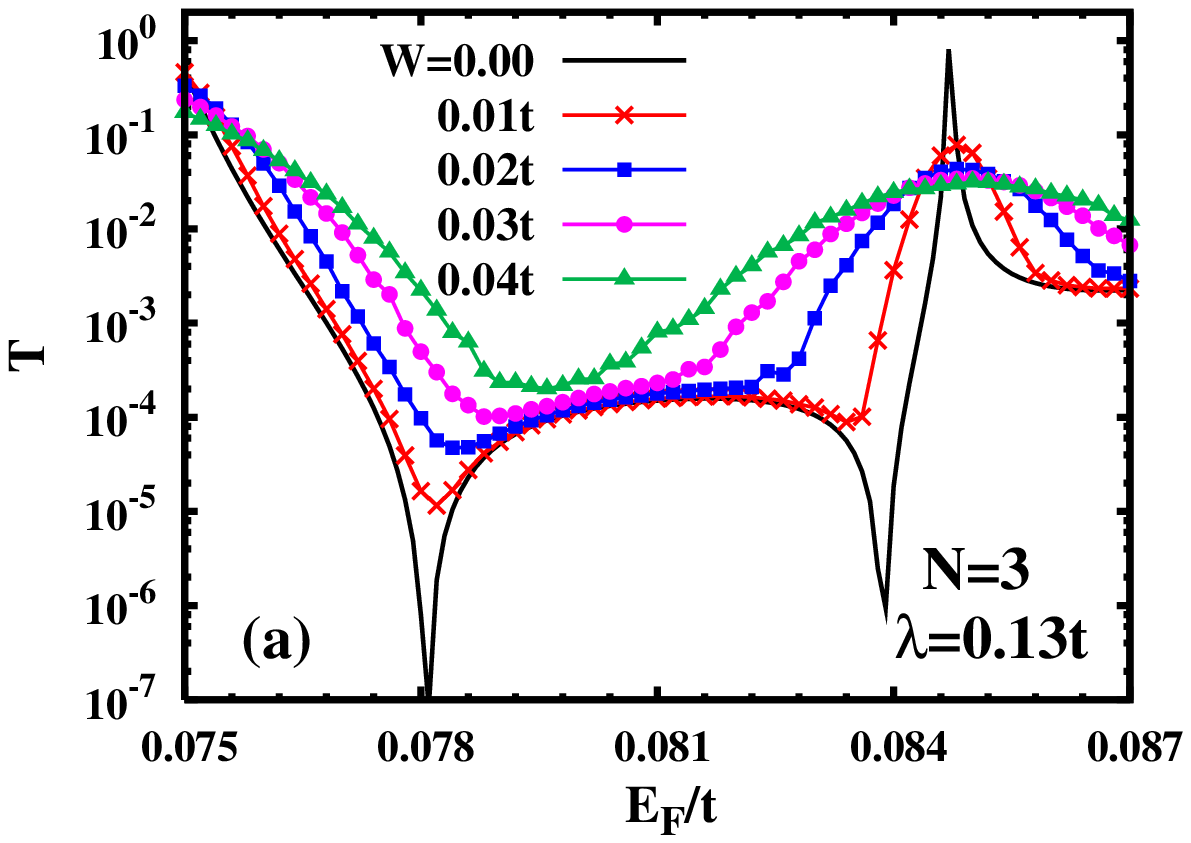}
\includegraphics[width=4.2cm]{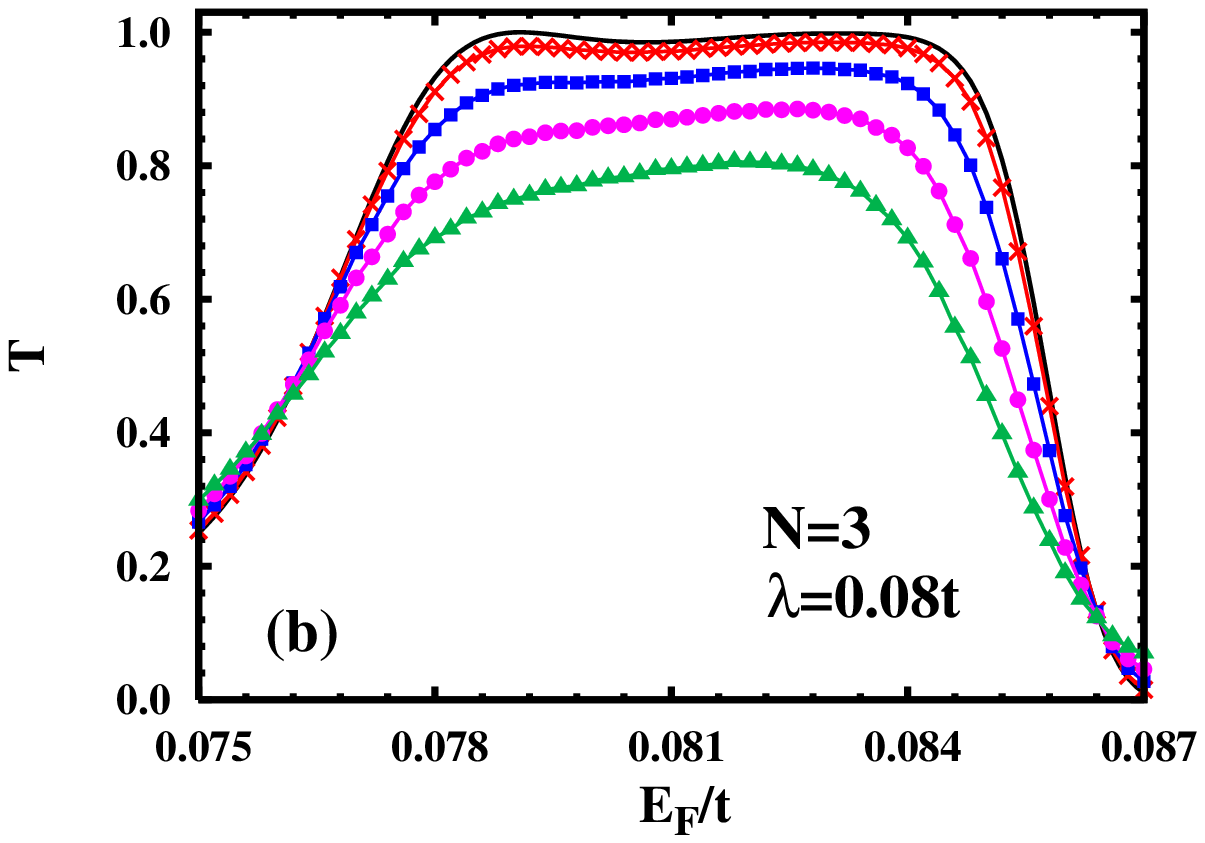}
\includegraphics[width=4.2cm]{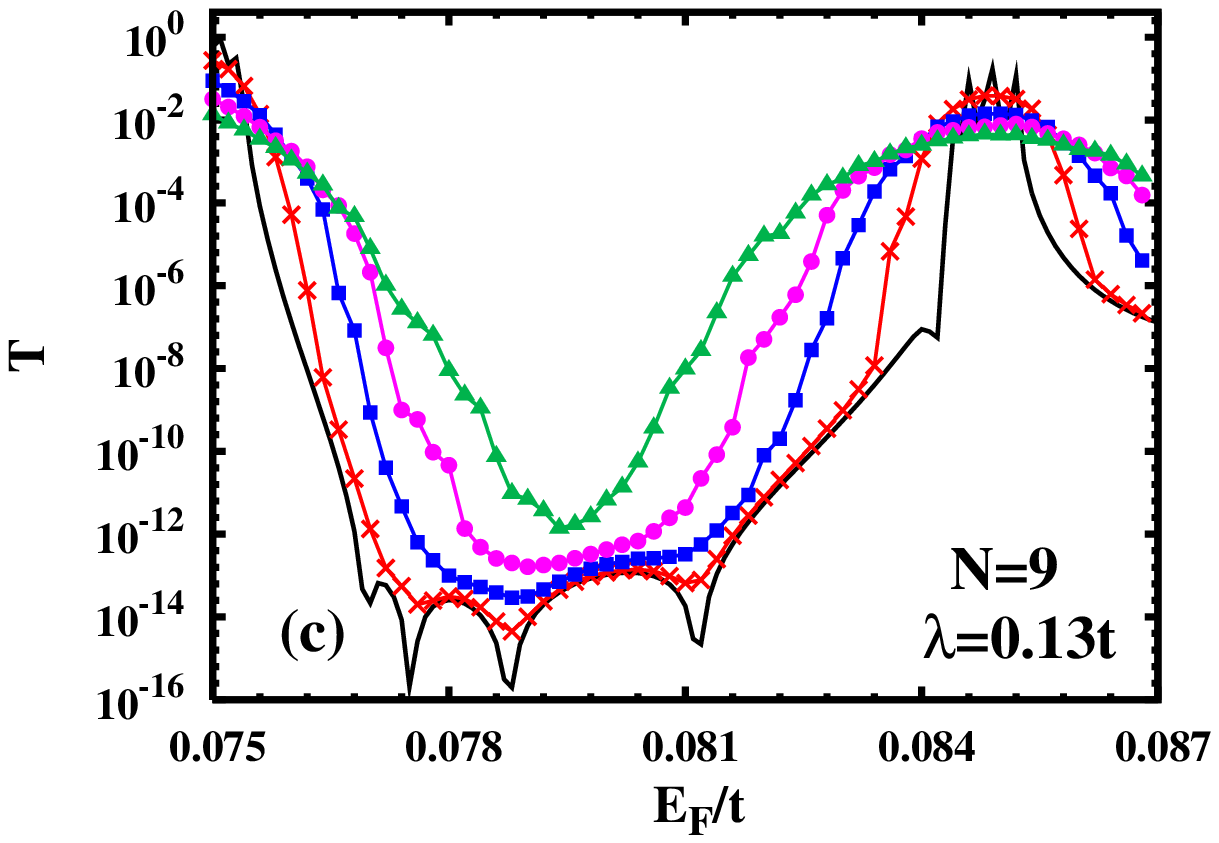}
\includegraphics[width=4.2cm]{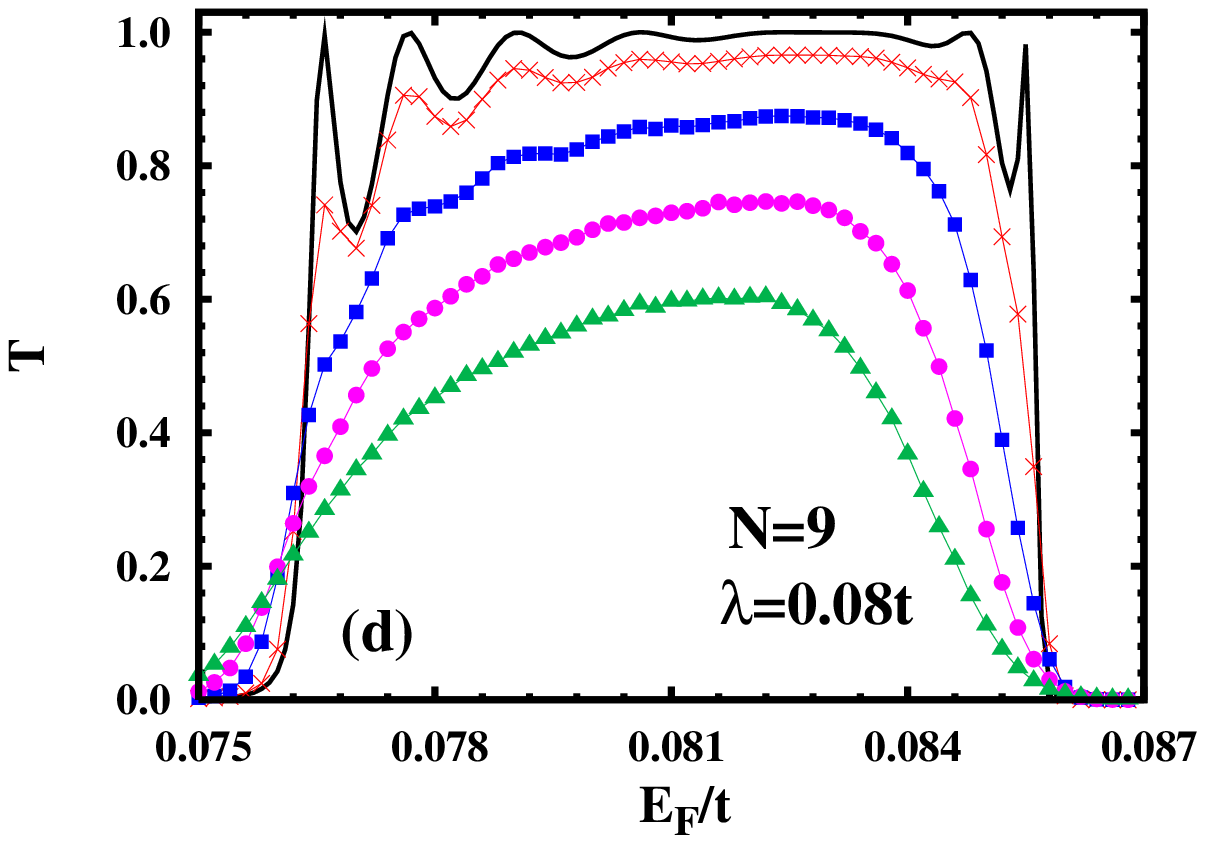}
\caption{(Color online)  Transmittances $T$ as a function of the
 Fermi energy of the leads for various Anderson disorder strengths $W$.
 For $N=3$, (a) $\lambda=0.13t$ (off current) and (b) $\lambda=0.08t$ (on current).
 For $N=9$, (c) $\lambda=0.13t$ (off current) and (d) $\lambda=0.08t$ (on current).}
\label{fig5}
\end{figure}

 Device disorders can destroy spin coherence. Thus, the 
 device performance may be spoiled severely due to the disorders. 
 Figure~5 shows the effects of the Anderson disorders 
 on electron transmission in (a) and (b) for $N=3$ and in (c) and (d) for $N=9$. 
 The transmission amplitudes were averaged over $1\times 10^4$ 
 configurations for $N=3$ and $3\times 10^4$ configurations for 
 $N=9$. 
 It is shown that the vanishingly small transmissions within the gaps 
 become larger as the disorder strength $W$ increases. 
 However, the transmission amplitudes are still smaller than 
 $T \lesssim 10^{-5}$ even for the stronger disorder $W=0.04t$, 
 which guarantees  a high performance of our SFETs. 
 This is clearly shown in Fig.~6. 
 The on/off current ratio decreases as the Anderson disorder strength 
 becomes stronger. 
 For $N=3$, relatively, a mild spoil of device performance occurs 
 because 
 the ratios are $I_\mathrm{on}/I_\mathrm{off}\simeq 1\times 
 10^4$ for $W=0.0$ and $I_\mathrm{on}/I_\mathrm{off}\simeq 
 0.89 \times 10^3$ for $W=0.04t$. 
 For $N=9$, 
 the quantum wire without the disorder 
 has a extremely large on/off current 
 ratio $I_\mathrm{on}/I_\mathrm{off}\simeq 2.7 \times 10^{12}$. 
 The strong disorder $W=0.04t$ gives rise to strong spin decoherence and 
 the on/off current ratio becomes $I_\mathrm{on}/I_\mathrm{off}\simeq 
 3.0 \times 10^5$. 
 However, this shows that the device performance is still good 
 enough for a SFET. 
 As a result, 
 the quantum wires with the spin-dependent modulations 
 can provide a high performance SFET even though the strong disorder 
 destroys spin coherence severely. 

\begin{figure}
\centering
\includegraphics[width=7cm]{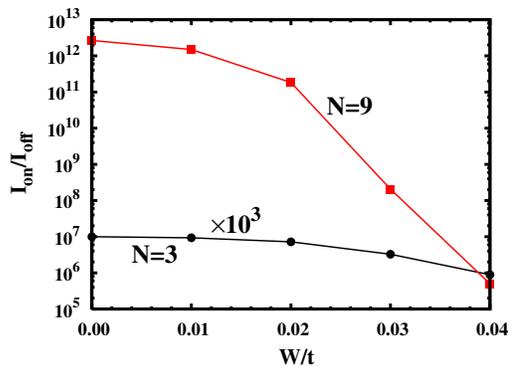}
\caption{(Color online) On-to-off current ratios
 $I_\mathrm{on}/I_\mathrm{off}$ as a function of
 the Anderson disorder strength $W$ for $N=3$ and $N=9$.} \label{fig6}
\end{figure}

\section{Summary}
 
 In summary, we have proposed the quantum wires with periodic local Rashba 
 SOCs as a spin transistor. 
 The device shows a good device performance 
 as a SEFT with extremely large on/off current ratios, 
 although strong device disorders destroy spin coherence. 
 
\begin{acknowledgements}
 
This work was supported by the Natural Science Foundation of China 
under Grant No.\ 10725417, the National Basic Research Program of 
China under Grant No.\ 2006CB922005 and the Knowledge Innovation 
Project of Chinese Academy of Sciences. S.Y.C. acknowledges the support 
from the NSFC under Grant No.\ 10874252 and the National Science 
Foundation Project of CQ CSTC. 
 
\end{acknowledgements}

\end{document}